\documentclass[preprint,showpacs,preprintnumbers,amsmath,amssymb]{revtex4}

\usepackage{graphicx}
\usepackage{dcolumn}
\usepackage{bm}

\def\VEC#1{\mbox{\boldmath $#1$}}

\begin{document}

\preprint{APS/123-QED}

\title{Propagation of Electromagnetic Waves
in Resistive Pair Plasma and Causal Relativistic 
Magnetohydrodynamics}

\author{Shinji Koide}
\affiliation{%
Faculty of Science, Kumamoto University,
2-39-1, Kurokami, Kumamoto 860-8555, Japan 
}%


\date{\today}

\begin{abstract}
We investigate the propagation of electromagnetic
waves in resistive e$^\pm$ pair plasmas using a one-fluid
theory derived from the relativistic two-fluid
equations. When the resistivity normalized by the electron/positron inertia
variable exceeds a critical value, the dispersion relation for
electromagnetic waves shows that the group velocity 
is larger than the light speed in vacuum.
However, in such a case, it also is found that the plasma parameter is less than unity: 
that is, the electron--positron pair medium no longer can be treated as plasma.
Thus the simple two-fluid approximation is invalid. This confirms that 
superluminal propagation of electromagnetic wave is forbidden
in a plasma ----- a conclusion
consistent with the relativistic principle of causality.
As an alternative, we propose a new set of equations
for ``causal relativistic magnetohydrodynamics", which both have non-zero
resistivity and yet are consistent with the causality principle.

\end{abstract}

\pacs{52.27.Ny, 52.30.Cv, 52.30.Ex, 52.35.Hr}
\maketitle

\section{Introduction}

Relativistic magnetohydrodynamic (RMHD) numerical simulations
have been performed by a number of groups recently \cite{koide98,koide99,koide00,koide02,
koide03,koide04,koide06,mizuno04,komissarov04,komissarov07,gammie03,mckinney04,mckinney06}.
These numerical simulations revealed many important, interesting features
of relativistic plasmas, especially around rotating black holes in
active galactic nuclei (AGNs), microquasars, and gamma-ray bursts.
Regarding energy extraction from a rotating black hole,
the magnetohydrodynamic (MHD) Penrose process has been confirmed \cite{koide02,koide03}, 
and long-term simulations of relativistic jet formation
around a rotating black hole have been performed \cite{mckinney06}.
All of these RMHD simulations
were restricted by the ideal MHD condition, where
electric resistivity is zero. 
In spite of recent significant advancements in ideal RMHD simulations,
one with finite resistivity (resistive RMHD) 
have not been performed seriously except in a few cases 
(e.g. \cite{watanabe06,komissarov07b}).
This is reasonable because there has been concern that the inclusion of finite 
resistivity in the RMHD equations destroys their causality.
In fact, the group velocity of electromagnetic waves 
derived mathematically from the resistive RMHD equations 
is larger than the light speed in vacuum.
This raises the possibility of superluminal communication, 
which is contradictory to the relativistic principle of causality.
The main purpose of this paper is to clarify and rectify this problem.

To fix this problem, we must reconsider the resistive RMHD
equations. Such a task was first performed
by \cite{ardavan76} using the Vlasov--Boltzmann equation
for a pulsar magnetosphere. It yielded a relativistic
version of the generalized Ohm's law and a new
condition for the validity of the MHD approximation
for a pulsar magnetosphere 
(where the Lorentz factor is much larger than unity).
A more generalized treatment, which included 
annihilation of electrons and positrons, radiation, Compton scattering,
and pair photoproduction was formulated by \cite{blackman93}
and \cite{gedalin96}.
Reconsideration of ideal MHD in a neutral cold plasma based on two-fluid
approximation was presented by \cite{melatos96}, who investigated the
conditions under which the MHD approximation
breaks down. For investigation of black hole magnetospheres,
\cite{khanna98} formulated the general relativistic version
of the two-fluid approximation in the Kerr metric.
An even more generalized version in a time-varying space-time 
was derived by \cite{meier04} from
the general relativistic Vlasov--Boltzmann equation.

In this present paper, we derive the one-fluid equations of an electron--positron
(pair) plasma based on the two-fluid equations with a new definition
of variable averaging for the two fluids (section \ref{sec2}).
In section \ref{sec3}, we derive the dispersion relation for electromagnetic
waves in uniform, unmagnetized and magnetized pair plasmas. 
We then examine the situation where the group
velocity of electromagnetic waves in the resistive plasma is 
larger than the light speed in vacuum and  
show that this condition cannot be realized in a plasma
whose plasma parameter is larger than unity.
In section \ref{sec4}, we propose a simple set of resistive
RMHD equations, which are consistent with the principle of causality.
In section \ref{sec5}, we discuss phenomena with respect to
the superluminal propagation of wave packets ----- phenomena that cannot
be avoided when the RMHD equations are acausal.
Finally, our summary and discussion are presented in section \ref{sec6}.

\section{Relativistic two--fluid model of pair plasma 
\label{sec2}}

To provide a solid base for resistive RMHD, we begin with
a relativistic two-fluid model of a pair plasma
in the Minkowski space-time $(x^0,x^1,x^2,x^3)=(t,x,y,z)$, where
the line element is given by
$ds^2=-(dx^0)^2+(dx^1)^2+(dx^2)^2+(dx^3)^2=\eta_{\mu\nu}dx^\mu dx^\nu$.
Throughout this paper (except for one paragraph in section \ref{sec3}), 
we use units in which the light speed, the dielectric constant,
and the magnetic permeability in vacuum all are unity: 
$c=1$, $\epsilon_0=1$, $\mu_0 = 1$.
The relativistic equations of the electron and positron
fluids are given as follows (e.g., \cite{misner70,weinberg72}):
\begin{eqnarray}
\partial_\mu (n_\pm U_\pm^\mu) &=& 0 , \label{4formnum} \\
\partial_\mu (h_\pm U_\pm^\mu U_\pm^\nu)  &=&
-\partial^\nu p_\pm \pm e n_\pm \eta^{\nu\sigma} U_\pm^\mu F_{\sigma\mu}
\pm R^\nu , \\
\partial_\mu \hspace{0.3em} ^*F^{\mu\nu} &=& 0 , \\
\partial_\mu F^{\mu\nu} & = & J^\nu , \label{4formmaxwel}
\end{eqnarray}
where a variable with subscript, plus (+) or minus (--),
is that of the positron and electron fluid, respectively,
$n_\pm$ is the proper particle number density, $U_\pm^\mu$ is the four-velocity,
$e$ is the electric charge
of positron, $p_\pm$ is the proper pressure, $h_\pm$ is the relativistic
enthalpy density, 
$F_{\mu\nu}$ is the electromagnetic field tensor, 
$\hspace{0.3em} ^*F^{\mu\nu}$ is the dual tensor density of $F_{\mu\nu}$,
$R^\mu$ is the frictional four-force density between the electron and
positron fluids, and $J^\mu$ is the four-current density. 
We often will write a set of the spacial components of the four-vector using a bold italic
font, e.g., $\VEC{U}_\pm = (U^1_\pm,U^2_\pm,U^3_\pm)$, $\VEC{J} = (J^1,J^2,J^3)$.
Here we assume that the electron/positron fluids are heated only
by Ohmic heating and neglect pair creation and annihilation.
We also neglect radiation and quantum effects.

We further define the Lorentz factor $\gamma_\pm = U^0_\pm$, the three-velocity
$V^i_\pm=U^i_\pm/\gamma_\pm$, the electric field $E_i=F^{0i}$, the magnetic flux
density $B_i=\frac{1}{2} \sum_{jk} \epsilon_{ijk} F^{jk}$
($\epsilon_{ijk}$ is the Levi--Civita tensor), and the electric charge density
$\rho_{\rm e} = J^0$. 
Here, the alphabetic index ($i,j,k$) runs from 1 to 3.
Using the above relativistic equations (\ref{4formnum})--(\ref{4formmaxwel}), 
we obtain the vector form of the relativistic two-fluid equations,
\begin{eqnarray}
\frac{\partial}{\partial t} (\gamma_\pm n_\pm) +
\nabla \cdot (n_\pm \VEC{U}_\pm) &=& 0 , \label{twofluidnum} \\
\frac{\partial}{\partial t} (h_\pm \VEC{U}_\pm)
+ \nabla \cdot (h_\pm \VEC{U}_\pm \VEC{U}_\pm)
\hspace{-0.2cm} &=& \hspace{-0.2cm} 
-\nabla p_\pm \pm e \gamma_\pm n_\pm (\VEC{E} + \VEC{V}_\pm \times \VEC{B})
\pm \VEC{R}, \\
\frac{\partial}{\partial t} (\gamma_\pm^2 h_\pm-p_\pm)
+ \nabla \cdot (\gamma_\pm h_\pm \VEC{U}_\pm ) 
& = & \pm e n_\pm \VEC{U} \cdot \VEC{E} \pm R^0  , \\
\nabla \cdot \VEC{E} = \rho_{\rm e} &=& e(\gamma_+ n_+ - \gamma_- n_-)  , \\
\nabla \cdot \VEC{B} &=& 0 , \\
\frac{\partial \VEC{B}}{\partial t} &=& - \nabla \times \VEC{E} , \\
\frac{\partial \VEC{E}}{\partial t} + \VEC{J} &=& \nabla \times \VEC{B} .
\label{twofluidmaxwel}
\end{eqnarray}
The frictional four-force density between electrons and positrons is
(equation (\ref{frc4frc2}) in Appendix \ref{appena}),
\begin{equation}
R^\mu=-\frac{m\nu_{\rm ee}}{n}
(n_- \gamma_-^\prime n_+ U_+^\mu - n_+ \gamma_+^\prime n_- U_-^\mu)
+ \frac{n_+ U_+^\mu + n_- U_-^\mu}{n_+ \gamma_+^\prime + n_- \gamma_-^\prime} 
 R^{0\prime} ,
\label{friction4frc}
\end{equation} 
where $\nu_{\rm ee}$ is the electron--positron Coulomb collision frequency,
$m$ is the mass of an electron/positron particle,
the variables with primes are physical quantity observed in the
center-of-mass frame of the two fluids, and $R^{0\prime}$ is the energy gain rate of the positron
fluid due to the friction with electron fluid in the center-of-mass frame.
Note that the variables observed in the center-of-mass frame are 
proper variables (see Appendix \ref{appena}).
If we assume that relative velocity of the positron and electron fluids is
much smaller than thermal velocity of the fluids, the collision
frequency $\nu_{\rm ee}$ is proportional to the relative velocity
of the two fluids and its coefficient depends only on 
temperature and density of the two fluids.
On the other hand, if the relative velocity of the positron and electron fluids is 
relativistic, then the coefficient also depends on their relative velocity.
Note also that when the relative velocity of the electron and positron fluids
is nonrelativistic, $\gamma'_\pm \rightarrow 1$, while in the case of relativistic
relative velocity, $\gamma'_\pm > 1$.
Through this paper, we usually assume that the relative velocity of the two fluids
is smaller than the sound velocity of the plasma (which is  nonrelativistic).

To derive the one-fluid equations of a pair plasma, we define average and difference
variables as follows:
\begin{eqnarray}
n &=& \frac{n_+ + n_-}{2}, \label{avenum} \\
\rho &=& 2mn, \label{avemas} \\
U^\mu &=& \frac{n_+ U_+^\mu + n_- U_-^\mu}{2n}  ,
\label{ave4vel} \\
J^\mu &=& e(n_+ U_+^\mu - n_- U_-^\mu) ,
\label{ave4cur}  \\
\tilde{h}&=& n^2 \left ( \frac{h_+}{n_+^2} + \frac{h_-}{n_-^2} \right ), \\
\Delta \tilde{h}&=&n^2 \left ( \frac{h_+}{n_+^2} - \frac{h_-}{n_-^2}\right ). 
\end{eqnarray}
From the relativistic two-fluid model of the 
pair plasma (\ref{twofluidnum})--(\ref{friction4frc}), 
we then can obtain the one-fluid equations of the pair plasma,
\begin{eqnarray}
&& \frac{\partial}{\partial t} (\gamma \rho) +
\nabla \cdot (\rho \VEC{U}) = 0  , \label{rmhdmass} \\
&& \frac{\partial}{\partial t} \left [ \tilde{h} \left ( \gamma \VEC{U} + 
\frac{1}{(2ne)^2} \rho_{\rm e} \VEC{J} \right ) +
\frac{\Delta \tilde{h}}{2en} \left ( \gamma \VEC{J} + \rho_{\rm e} \VEC{U} \right ) 
\right ]
+ \nabla \cdot \left [ \tilde{h} \left ( 
\VEC{UU} + \frac{1}{(2en)^2} \VEC{JJ} \right ) 
+ \frac{\Delta \tilde{h}}{2en} (\VEC{UJ} + \VEC{JU}) \right ]  \nonumber \\
&& = 
-\nabla p + \rho_{\rm e} \VEC{E} + \VEC{J} \times \VEC{B} , \label{rmhdmomentum} \\
&& \VEC{E} + \VEC{V} \times \VEC{B} + \frac{1}{2en} \nabla (p_- - p_+)
 -  \eta \frac{\gamma'}{\gamma} \left [\VEC{J} - \frac{1 + \Theta}{\gamma^{\prime 2}} 
(\gamma \rho_{\rm e} - \VEC{J} \cdot \VEC{U} ) \VEC{U} \right ]  \nonumber \\
&& = \frac{1}{4ne^2\gamma} \left [ \frac{\partial}{\partial t} \left ( 
\frac{\tilde{h}}{n} ( \gamma \VEC{J} + \rho_{\rm e} \VEC{U} ) 
+ \Delta \tilde{h} \left \{ \gamma \VEC{U} 
+ \frac{1}{(2ne)^2} \rho_{\rm e} \VEC{J} \right \} \right )
\right . \nonumber \\
&& +  \left . 
\nabla \cdot \left \{ \frac{\tilde{h}}{n} (\VEC{UJ} + \VEC{JU}) 
+ 2 n^2 e \Delta \tilde{h} \left ( \VEC{UU} + \frac{1}{(2en)^2} \VEC{JJ} \right )\right \}
 \right ] \label{rmhdohm}, \\
&& \frac{\partial}{\partial t} \left [ \tilde{h} \left ( \gamma^2 + \frac{\rho_{\rm e}^2}{(2ne)^2} \right ) 
+ \Delta \tilde{h} \frac{\gamma}{ne}  \rho_{\rm e} -p \right ] 
+ \nabla \cdot \left [ \tilde{h} \left ( 
\gamma \VEC{U} + \frac{1}{(2ne)^2} \rho_{\rm e} \VEC{J} \right ) +
\frac{\Delta \tilde{h}}{2ne} \left ( \gamma \VEC{J} + \rho_{\rm e} \VEC{U} \right ) 
\right ]   \nonumber = \VEC{J} \cdot \VEC{E} , \nonumber \\
&& 
\end{eqnarray}
\begin{eqnarray}
\nabla \cdot \VEC{E} &=& \rho_{\rm e} , \\
\nabla \cdot \VEC{B} &=& 0 , \\
\frac{\partial \VEC{B}}{\partial t} &=& - \nabla \times \VEC{E} , \\
\frac{\partial \VEC{E}}{\partial t} + \VEC{J} &=& \nabla \times \VEC{B}  ,
\label{rmhdampere}
\end{eqnarray}
where $\eta = m \nu_{\rm ee}/(n e^2)$ is the electric resistivity and
$\Theta$ is the equipartition factor of the thermal energy due to
the friction between the electron and positron fluids
given in Appendix \ref{appena}.

Equation (\ref{rmhdohm}) corresponds to the generalized Ohm's law.
The classical formulation of the resistivity $\eta$ 
of a non-magnetized nonrelativistic plasma (due to 
Coulomb collision) is used, and the formulation for an electron--proton
plasma is confirmed as an appropriate expression for resistivity
of weakly magnetized nonrelativistic plasma in laboratory experiments
(e.g., in a ``Tokamak" thermonuclear fusion device \cite{bellan06}).
Note that the Hall effect disappears in a pair plasma.

\section{Dispersion relation for electromagnetic waves in resistive plasma
\label{sec3}}

In this section, we derive the dispersion relation for electromagnetic waves in a
pair plasmas using a linear analysis of equations
(\ref{rmhdmass})--(\ref{rmhdampere}). 
First, we assume that the background plasma is at rest, uniform, and non-magnetized:
$\rho=\rho_0$ ($n=n_0$), $\VEC{V}=\VEC{0}$, $p=p_0$ ($p_+=p_-=p_0/2$), 
$\tilde{h}=h_0$, $\VEC{B}=\VEC{E}=\VEC{0}$.
Perturbations due to the electromagnetic waves are so small that
the plasma motion is non-relativistic. 
Then we have the following linearized equations with respect to the perturbations,
$\rho_1=\rho - \rho_0$ ($n_1 = n - n_0$), $\VEC{V}_1=\VEC{V}$, $p_1=p-p_0$,
$\VEC{B}_1=\VEC{B}$, $\VEC{E}_1=\VEC{E}$, $\VEC{J}_1=\VEC{J}$,
$\rho_{\rm e 1} = \rho_{\rm e}$,
$h_1=\tilde{h}-h_0$,
\begin{eqnarray}
\frac{\partial}{\partial t} \rho_1 +
\nabla \cdot (\rho_0 \VEC{V}_1) &=& 0 ,\\
h_0 \frac{\partial \VEC{V}_1}{\partial t} &=& - \nabla p_1 ,\\
\VEC{E}_1 - \eta \VEC{J}_1 &=& \kappa \frac{\partial \VEC{J}_1}{\partial t} , 
\label{linearohm} \\
\frac{\partial}{\partial t} (h_1 - p_1) + \nabla \cdot (h \VEC{V}_1) &=& 0 , \\
\nabla \cdot \VEC{E}_1 &=& \rho_{\rm e 1} , \\
\nabla \cdot \VEC{B}_1 &=& 0 , \\
\frac{\partial \VEC{B}_1}{\partial t} &=& - \nabla \times \VEC{E_1} , \\
\frac{\partial \VEC{E}_1}{\partial t} + \VEC{J}_1 &=& \nabla \times \VEC{B}_1 ,
\end{eqnarray}
where $\kappa = h_0/(2n_0e)^2$, and we assume $p_+^1 \approx p_-^1$.
These linearized equations do not depend on the equipartition fraction 
of frictionally thermalized energy, $\theta$ ($0 \le \theta \le 1$).
When we consider transverse modes of the linearized equations, 
$\VEC{k} \cdot \VEC{V_1} = 0$, the dispersion relation for 
the electromagnetic waves can be written as
\begin{equation}
( k^2 - \omega^2) \left ( 1 - \frac{i\omega}{2\nu_{\rm ee}'} \right )
= i\frac{\omega}{\eta}  ,
\label{thedisprel0}
\end{equation}
where $\VEC{k}$ is the wave number vector, $\omega$ the frequency of 
the electromagnetic wave, and
$\nu_{\rm ee}' = \eta/(2\kappa) = (mn_0/h_0) \nu_{\rm ee} $.
Here we also obtain $\rho_1 = 0$, $p_1 = 0$, $\rho_{\rm e 1} = 0$.

\begin{figure}
\includegraphics[width=0.85\textwidth]{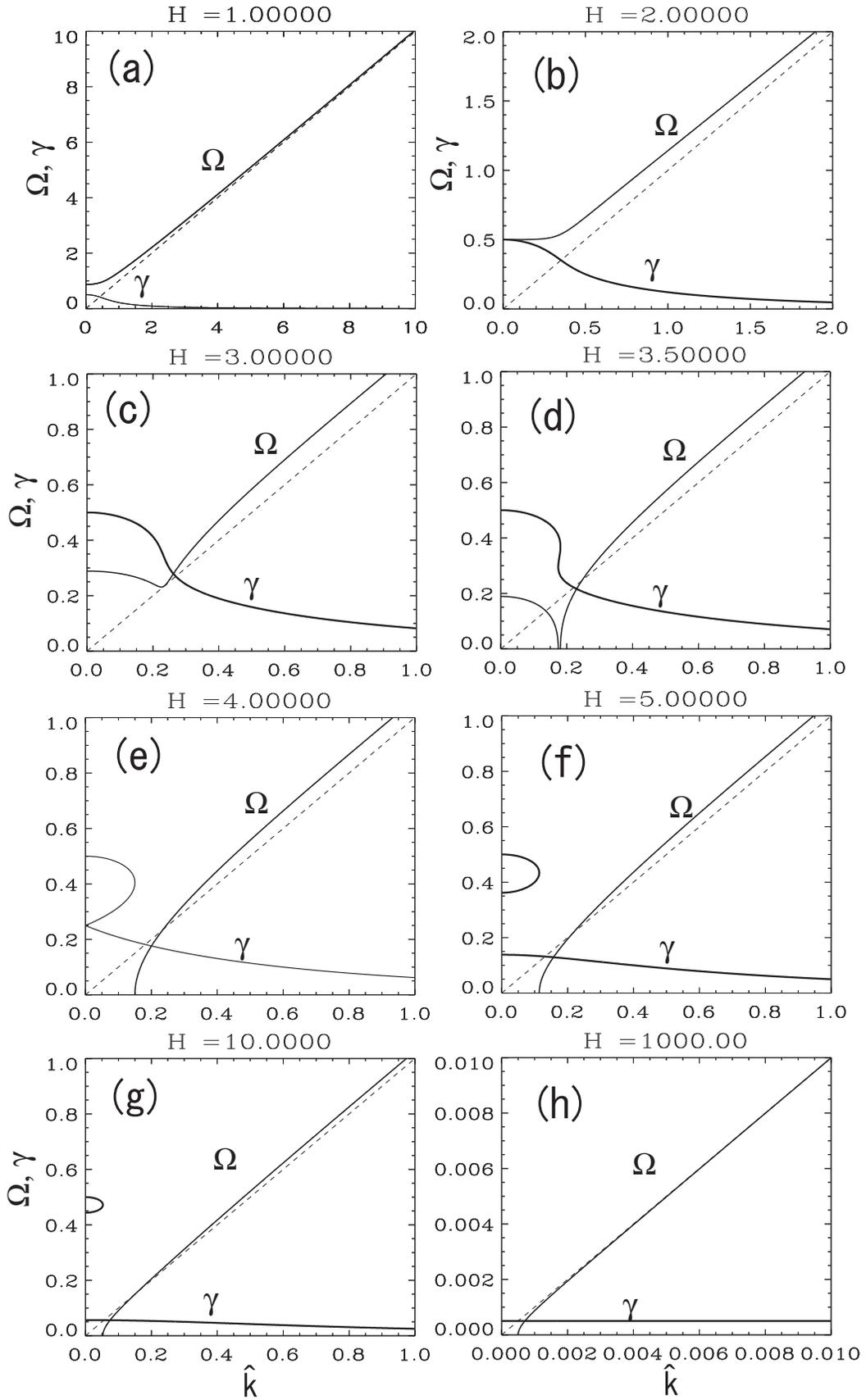}
\caption{
Dispersion relation for electromagnetic waves in the
resistive pair plasma for various $H$.
The dotted line shows the dispersion relation for electromagnetic waves
in vacuum.
\label{disprels8}
}
\end{figure}

\begin{figure}
\includegraphics[width=0.5\textwidth]{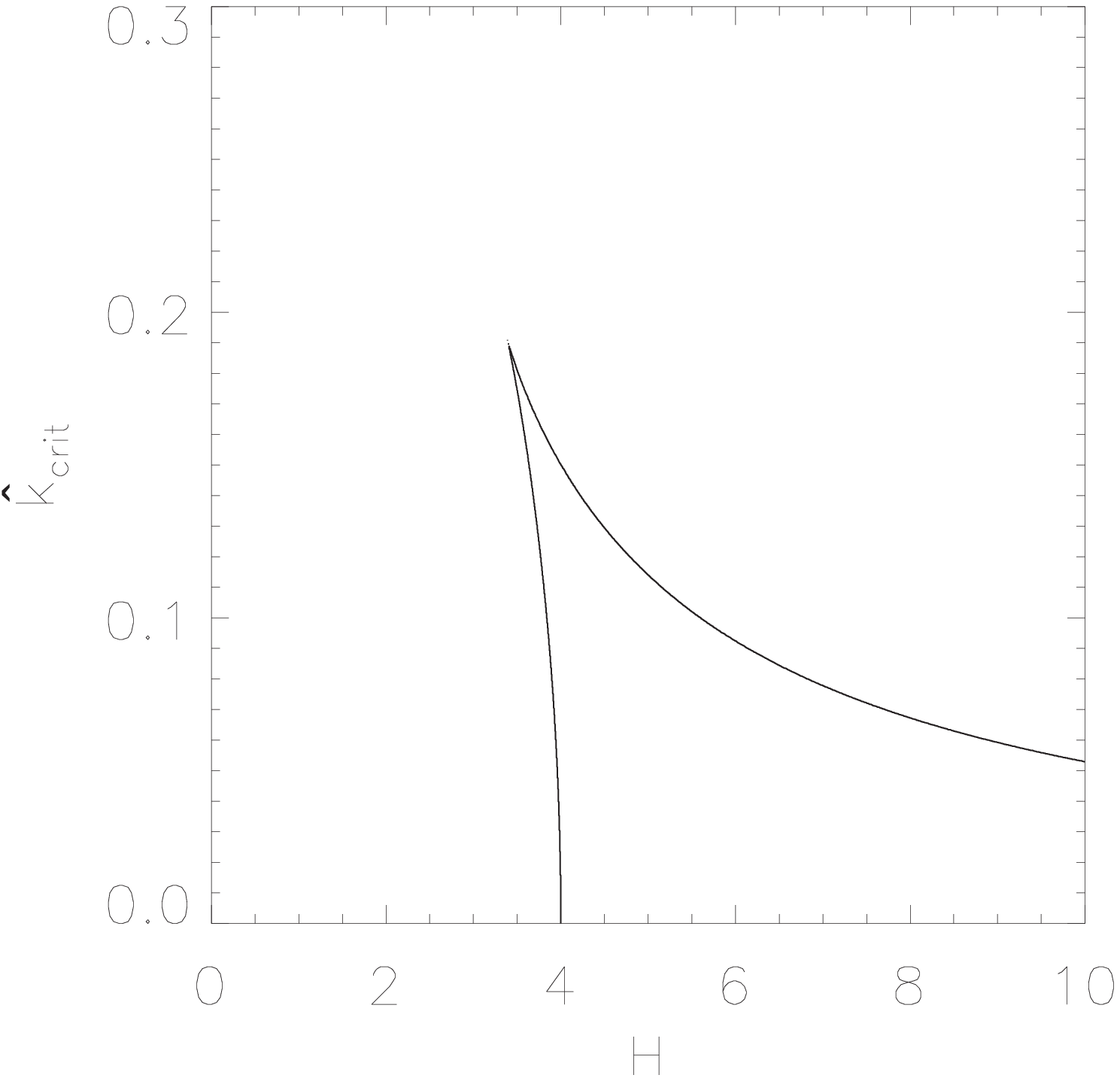}
\caption{
Dependence of $\hat{k}_{\rm crit}$ on $H$. At $\hat{k} = \hat{k}_{\rm crit}+0$, 
$\partial \Omega/\partial \hat{k}$ becomes infinity for each $H$.
\label{calkcri}
}
\end{figure}

When we normalize the variables as $\hat{\omega} = \omega/ (2 \nu_{\rm ee}')$
and $\hat{k} = k/ (2 \nu_{\rm ee}')$, the dispersion relation (\ref{thedisprel0}) 
becomes
\begin{equation}
H(\hat{\omega}^2 - \hat{k}^2) (1 - i \hat{\omega}) + i \hat{\omega} = 0,
\label{dispcomlex}
\end{equation}
where $H = 2 \eta \nu_{\rm ee}' = \eta^2/\kappa$.
Note that the parameter $H$ is related to 
the Coulomb collision frequency $\nu_{\rm ee}=\eta n_0 e^2/m$ and
the electron plasma frequency $\omega_{\rm pe} = (n_0e^2/m)^{1/2}$ as
$H=2(mn_0/h_0) (\nu_{\rm ee}/\omega_{\rm pe})^2$.
Setting $\hat{\omega} = \Omega - i \gamma$ ($\Omega \ge 0, \gamma \in {\bf R}$),
we obtain the dispersion relation with respect to the real frequency $\Omega$
and the damping rate $\gamma$,
\begin{equation}
\gamma^3 - \gamma^2 + \frac{1}{4} \left ( 1 + \hat{k}^2 + \frac{1}{H} \right )\gamma
- \frac{1}{8H} = 0   ,
\label{disprealgamma}
\end{equation}
and 
\begin{equation}
\Omega^2 ( 4 \Omega^2 -C )^2 - \frac{1}{27} (C^3 + F^2) = 0   ,
\label{disprealomega}
\end{equation}
where $C = 3 (\hat{k}^2 + 1/H)-1$, $F = 9 [\hat{k}^2 - 1/(2H)]+1$.
In this section and Appendix \ref{appenb}, we use $\gamma$ to denote
the damping rate.
We also have the relation between $\Omega$ and $\gamma$, 
\begin{equation}
\Omega^2 = 3\gamma^2 - 2\gamma + \hat{k}^2 + \frac{1}{H} .
\label{omega2gamma0}
\end{equation}
The dispersion relations with various $H$ are shown in Fig. \ref{disprels8}.
Note that the determinant of the cubic equation (\ref{disprealgamma}) 
with respect to $\gamma$
is $D_\gamma = - (C^3 + F^2)/(9 \times 36^2)$. If $\gamma$ has 
three different real solutions, $D_\gamma > 0$, i.e.,
$\Omega^2 (4 \Omega^2 -C)^2 = (1/27) (C^3 + F^2) < 0$, then $\Omega$ has
no real solution. Therefore, we have to consider range of the single
$\gamma$ solution, $D_\gamma \le 0$.
This range is given by $\hat{k} > \hat{k}_{\rm crit}$, where
the critical wave number $\hat{k}_{\rm crit}$ is defined by $\Omega = 0$ 
if a solution of $\Omega=0$ exists and $\hat{k}_{\rm crit} = 0$
if there is no solution (see Fig. \ref{disprels8}).
These results clearly show that 
the group velocity $v_{\rm g} = \partial \Omega/\partial \hat{k}$ is larger 
than one when $H \agt 3$; that is, superluminal wave packet propagation is possible 
(see Appendix \ref{appenc}).
When $H \ge 3.5$, there are points of $\Omega = 0$ at 
$\hat{k} = \hat{k}_{\rm crit} >0$.
Figure \ref{calkcri} shows the value of $\hat{k}_{\rm crit}$ for each $H$ case.
The group velocity $\partial \Omega/\partial \hat{k}$ is infinity 
at $\hat{k} \rightarrow \hat{k}_{\rm crit} + 0$.
On the other hand, in the cases of $H=1$ and $H=2$ 
(see Fig. \ref{disprels8}(a) and (b)), 
the derivative of $\Omega$ with respect to $\hat{k}$ increases monotonically, and 
$\partial \Omega/\partial k$ approaches unity when $\hat{k}$ becomes infinity 
($\lim_{\hat{k} \longrightarrow \infty} \partial \Omega/\partial \hat{k} = 1$);
that is, the gradient remains less than unity as long as $\hat{k}$
remains finite. 
Furthermore, we prove that $\partial \Omega/\partial \hat{k} <1$ 
when $H<1.5$ (see Appendix \ref{appenb}). We find that there is no possibility of 
superluminal propagation of electromagnetic wave in the case of $H<2$,
while $\partial \Omega/\partial k$ is larger than unity in a certain range
of $\hat{k}$ when $H \ge 3$. (A detailed investigation produces a more
strict condition on superluminal propagation of $H \ge 2.3$.)

We now show that matter composed of electrons and positrons 
with $H \ge 1$ cannot be treated as a plasma.
This means that superluminal propagation of electromagnetic waves
is not permitted when the medium is a plasma; and, when it is not,
the medium must be treated in a different manner. 
Note that in this paragraph only, we shall use the SI unit system.
The plasma parameter is given by
\begin{equation}
N_{\rm p} = n \lambda_{\rm D}^3
= \sqrt{\frac{(\epsilon_0 T)^3}{n e^6}}   ,
\end{equation}
where $T$ is the temperature of the electron/positron fluids and
$\lambda_{\rm D}$ is the Debye length, 
$\lambda_{\rm D} = \epsilon_0 T/(n e^2)$ \cite{bellan06}.
For a plasma, $N_{\rm p}$ is (much) larger than unity
because charged particles are bound to
each other when $N_{\rm p} < 1$.
The frequency of electron--positron Coulomb collisions can be written
as,
\begin{equation}
\nu_{\rm ee} = \frac{n e^4 \ln \Lambda}{6 \sqrt{3} \pi
\epsilon_0^2 \sqrt{m} T^{3/2}}
=\frac{n e^4}{\epsilon_0^2 \sqrt{m} T^{3/2}} \ln \Lambda',
\end{equation}
where 
$\ln \Lambda$ is the Coulomb logarithm and $\ln \Lambda' = \ln \Lambda
/(6 \sqrt{3} \pi) \sim 1/3$ \cite{bellan06}. Here we used $\ln \Lambda \sim 10$. 
Then we find that 
$H= 2(mn/h)(\nu_{\rm ee}/\omega_{\rm pe})^2< 2 (\nu_{\rm ee}/\omega_{\rm pe})^2 
=2(\ln \Lambda)^2 n e^6/(\epsilon_0 T)^3$.
Finally, we get the relation between $H$ and $N_{\rm p}$,
\begin{equation}
H N_{\rm p}^2 < 2 (\ln \Lambda')^2 \sim \frac{2}{9}.
\end{equation}
Therefore, when we consider a plasma (i.e., $N_{\rm p} > 1$), 
we find that $H<(\ln \Lambda')^2/N_{\rm p}^2 \alt 2/9 < 1$.
This clearly shows superluminal propagation of electromagnetic 
wave is not permitted  in a true plasma (usually, $N_{\rm p} \gg 1$).

In the above discussion we used the rough approximation $\ln \Lambda \sim 10$.
If $\ln \Lambda$ were greater than 100 separately with other variables, 
$H$ would become larger than
3 and then superluminal communication would become possible.
However, this situation can never be realized because there is strict relation
between $\Lambda$ and $N_{\rm p}$ as $\Lambda = 6 \pi N_{\rm p}$
(see \cite{bellan06}, page 24). So, if $\ln \Lambda$ becomes larger,
then $N_{\rm p}$ becomes much larger and $H$ decreases to a value much less
than unity.

Next, we discuss briefly the dispersion relation for electromagnetic waves 
in a uniformly magnetized pair plasma. We assume the background
plasma is the same as that of the previous unmagnetized case except
for a uniform magnetic field, $\VEC{B} = \VEC{B}_0 \ne \VEC{0}$.
Using the same procedure employed in the previous unmagnetized 
plasma case, we obtain the linearized equations,
\begin{eqnarray}
\VEC{E}_1 + \frac{i}{\omega h_0} (\VEC{J}_1 \times \VEC{B}_0)
\times \VEC{B}_0 &=& (\eta - i \kappa \omega) \VEC{J}_1, \label{magplalinele1} \\
\omega^2 \VEC{E}_1 + i \omega \VEC{J}_1 = k^2 \VEC{E}_1, \label{magplalincur1}
\end{eqnarray}
where we assume $\VEC{k} \cdot \VEC{V}_1 = 0$ 
and $\VEC{k} \cdot \VEC{E}_1 = 0$ to investigate transverse modes.
When we separate the perturbations of the electric field and current density into 
two components parallel and perpendicular to the background magnetic field
$\VEC{B}_0$,
\begin{eqnarray}
\VEC{E}_1 = \VEC{E}_\parallel + \VEC{E}_\perp, &
\VEC{E}_\parallel \parallel \VEC{B}_0, &
\VEC{E}_\perp \perp \VEC{B}_0, \\
\VEC{J}_1 = \VEC{J}_\parallel + \VEC{J}_\perp, &
\VEC{J}_\parallel \parallel \VEC{B}_0, & 
\VEC{J}_\perp \perp \VEC{B}_0, \\
\end{eqnarray}
equations (\ref{magplalinele1}) and (\ref{magplalincur1}) yield
\begin{eqnarray}
\VEC{E}_\perp - \frac{iB_0^2}{\omega h_0} \VEC{J}_\perp 
&=& (\eta - i \kappa \omega) \VEC{J}_\perp, \label{magplalinele2} \\
(k^2 - \omega^2) \VEC{E}_\perp &=& i\omega \VEC{J}_\perp, \\
\VEC{E}_\parallel &=& (\eta - i \kappa \omega) \VEC{J}_\parallel, \\
(k^2 - \omega^2) \VEC{E}_\parallel &=& i\omega \VEC{J}_\parallel. \\
\end{eqnarray}
Finally, we obtain the two dispersion relations,
\begin{eqnarray}
(k^2 - \omega^2) \left (\eta -i \kappa \omega + \frac{iB_0^2}{\omega h_0} 
\right ) &=& i\omega,
\label{dispmagpla1} \\
(k^2 - \omega^2) (\eta - i \kappa \omega) &=& i\omega .
\label{dispmagpla2}
\end{eqnarray}
Equations (\ref{dispmagpla1}) and (\ref{dispmagpla2}) are not
satisfied simultaneously when $B_0 \ne 0$. Equation (\ref{dispmagpla2})
is the same as that of the unmagnetized pair plasma case. Therefore,
we shall investigate the dispersion relation (\ref{dispmagpla1}).

\begin{figure}
\includegraphics[width=\textwidth]{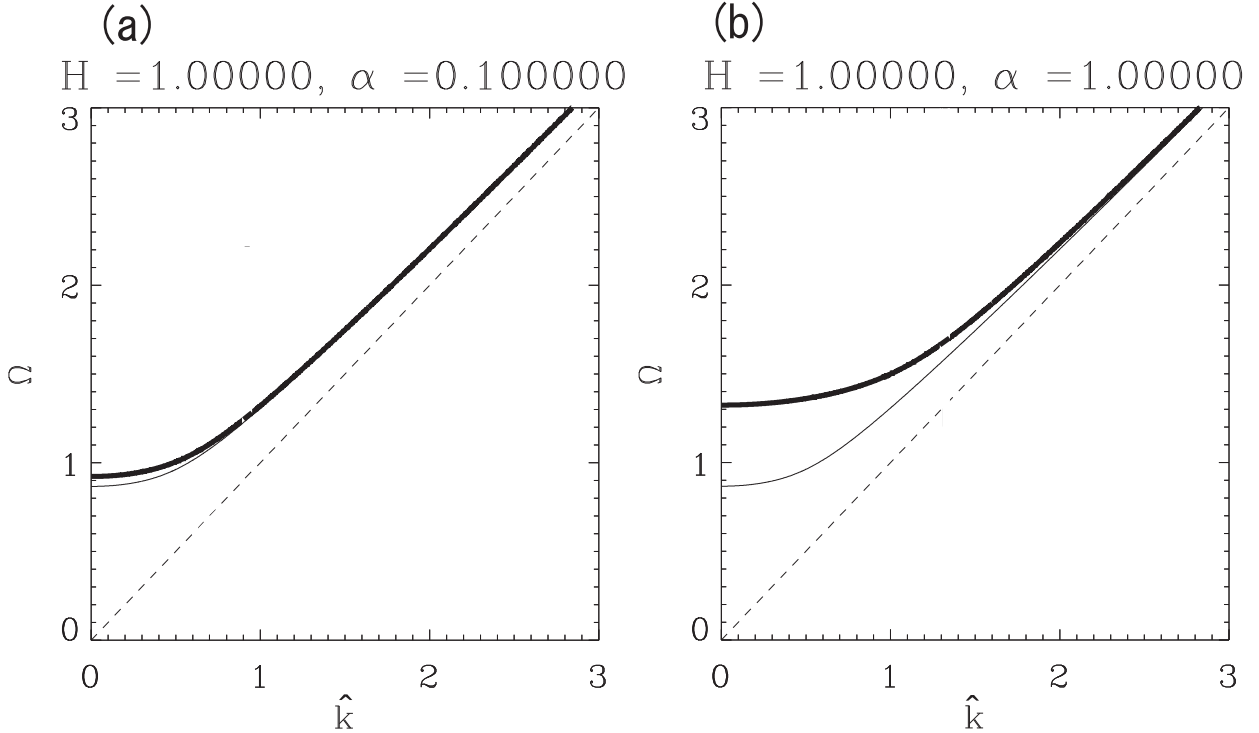}
\caption{The dispersion relation of electromagnetic waves in uniform,
magnetized pair plasma (thick solid lines) and in unmagnetized plasma
for comparison ($\alpha = 0$; thin solid lines).
(a) Sub-relativistically strong magnetic field case, $H=1$, $\alpha = 0.1$.
(b) Relativistically strong magnetic field case, $H=1$, $\alpha = 1$.
\label{dispmag}
}
\end{figure}

When we set $\hat{\omega} = \kappa \omega/\eta$, $\hat{k}=\kappa k/\eta$,
we have,
\begin{equation}
H(\hat{\omega}^2 - \hat{k}^2)(i\hat{\omega} + \hat{\omega}^2 -\alpha) = \omega^2,
\end{equation}
where $H=\eta^2/\kappa$ and $\alpha = B_0^2 \kappa/(h_0 \eta^2) = u_{\rm A}^2/H$
($u_{\rm A} \equiv B_0/\sqrt{h_0}$ is the Alfven four-velocity).
Setting $\hat{\omega} = \Omega -i \gamma$ ($\Omega \ge 0, \gamma \in {\bf R}$),
we obtain the dispersion relation for electromagnetic waves
in a magnetized pair plasma,
\begin{eqnarray}
H[(\Omega^2 -\gamma^2 -\hat{k}^2) (\Omega^2 - \gamma^2 - \alpha + \gamma)
+ 2 \Omega^2 \gamma (1-2 \gamma)] &=& \Omega^2 -\gamma^2, \\
H[(\Omega^2 -\gamma^2 -\hat{k}^2) (1 - 2 \gamma) - 2 \gamma
(\Omega^2 - \gamma^2 -\alpha + \gamma) ] &=& -2 \gamma, 
\end{eqnarray}
where we assume $\Omega \ne 0$.
Figure \ref{dispmag} shows the dispersion relation for $\Omega$ 
in the case  $H=1$ and $\alpha = 0.1, 1$.
The figure clearly shows the group velocity of electromagnetic
waves in a magnetized pair plasma is less than the light speed in vacuum
when $H=1$. A detailed investigation shows that this is true when $H<1$ 
as in the unmagnetized pair plasma case.
However, when $H > 4$, the group velocity is larger than the speed of light
for some ranges of $\hat{k}$. 

\section{Causal Resistive RMHD Equations
\label{sec4}}

In the above discussion, we derived the one-fluid equations 
(\ref{rmhdmass})--(\ref{rmhdampere}) from the two-fluid ones.
The one-fluid equations confirmed that the superluminal
propagation is forbidden in a two-component medium that has a
plasma whose plasma
parameter greater than unity. That is, the one-fluid equations
of a pair plasma (\ref{rmhdmass})--(\ref{rmhdampere}) are causal.
When we neglect the first term of the right hand side in Ohm's law
equation (\ref{rmhdohm}), which comes from the inertial effect of the
positron and electron, the term, $-i\omega/(2\nu'_{\rm ee})$, on the
left hand side of the dispersion relation (\ref{thedisprel0}) drops out.
In this case, the group velocity becomes $v_{\rm g} = \partial \omega/
\partial k= 2k(4k^2-\eta^{-2})^{-1/2} > 1$, which means the group 
velocity is greater than the speed of 
light (superluminal). As shown in Appendix \ref{appenc}, when 
the group velocity is larger than the speed of light, superluminal 
communication would become possible, allowing us to develop
a device that could send information into the past. 
However, such a device would destroy
the causality of time--ordered events and, therefore, should not be possible. 
This means that, in order to preserve causality, we cannot neglect the inertial
term of the electron and positron in Ohm's law (\ref{rmhdohm}).
Recently, several groups performed simulations of resistive RMHD
including Ohm's law without the electron/positron inertia effect
\cite{watanabe06,komissarov07b}. As shown in the above results, unfortunately,
all of these calculations are acausal. Here, we propose a set of causal 
resistive RMHD equations in a simple form.

For simplicity, assuming that $|\Delta \tilde{h}| \ll \tilde{h}$
and $p_+ \approx p_-$, we obtain the following equations,
\begin{eqnarray}
\frac{\partial}{\partial t} (\gamma \rho) + \nabla \cdot (\rho \VEC{U}) &=& 0
\label{causalrmhdmass},\\
\frac{\partial}{\partial t} \left [ \tilde{h} \left ( 
\gamma  \VEC{U} + \frac{\rho_{\rm e}}{(2ne)^2} \VEC{J} \right ) \right ]
+ \nabla \cdot \left [ \tilde{h} \left ( \VEC{UU} + \frac{1}{(2ne)^2} \VEC{JJ} \right ) 
\right ] &=& 
- \nabla p + \rho_{\rm e} \VEC{E} + \VEC{J} \times \VEC{B},
\label{causalrmhdmom}\\
\frac{\partial}{\partial t} \left [ \tilde{h} \left ( 
\gamma^2 + \frac{\rho_{\rm e}^2}{(2ne)^2} \right )  - p \right ] 
+ \nabla \cdot \left [ \tilde{h} \left (
\gamma \VEC{U} + \frac{\rho_{\rm e}}{(2ne)^2} \VEC{J} \right ) \right ] 
&=& \VEC{J} \cdot \VEC{E},
\label{causalrmhdenr} \\
\VEC{E} + \VEC{V} \times \VEC{B} - \frac{\eta}{\gamma} [\VEC{J} - \gamma^2
(\rho_{\rm e} - \VEC{V} \cdot \VEC{J}) (1 + \Theta ) \VEC{V}] && \nonumber \\
= \frac{1}{4ne\gamma} \left [ 
\frac{\partial}{\partial t} \left ( 
\frac{\tilde{h}}{ne} (\gamma \VEC{J} + \rho_{\rm e} \VEC{U} ) \right ) 
+ \nabla \cdot \left \{ 
\frac{\tilde{h}}{ne} (\VEC{UJ} + \VEC{JU})
\right \} \right ], && \label{causalrmhdohm} 
\end{eqnarray}
\begin{eqnarray}
\nabla \cdot \VEC{E} &=& \rho_{\rm e}, 
\label{causalrmhddive} \\
\nabla \cdot \VEC{B} &=& 0 ,\\
\frac{\partial \VEC{B}}{\partial t} &=& - \nabla \times \VEC{E},\\
\frac{\partial \VEC{E}}{\partial t} + \VEC{J} &=&  \nabla \times \VEC{B},
\label{causalrmhdampere}
\end{eqnarray}
where 
$\Theta = 2\theta (m/e)^2 (Q^2+W) \left /\left [\rho^2 - \left (mQ/e \right )^2 \right ] 
\right .$
(see Appendix \ref{appena}).
Here we have to assume $\gamma' \approx 1$ and $M=U_\nu U^\nu=-1$ 
in Appendix \ref{appena}, which means that the relative 
velocity of the electron fluid and positron fluid is nonrelativistic.
This condition also preserves $\gamma = 1/(1 - V^2)^{1/2}$.
In a pair plasma, we use $\theta=1$.

A covariant form for these one-component fluid equations
(\ref{causalrmhdmass})--(\ref{causalrmhdampere}) is as follows:
\begin{eqnarray}
\partial_\nu (\rho U^\nu) &=& 0, \\
\partial_\nu \left [ \tilde{h} \left ( U^\nu U^\mu + 
\frac{1}{(2ne)^2} J^\nu J^\mu \right ) \right ] &=&
-\partial^\mu p + J^\nu F_\nu^\mu, \\
U^\nu F_\nu^\mu - \eta [J^\mu + (U^\nu J_\nu) (1+\Theta) U^\mu]
&=& \frac{1}{4ne^2} \left [ \partial_\nu \left \{ \frac{\tilde{h}}{n}
\left ( U^\nu J^\mu + J^\nu U^\mu \right ) \right \} \right ], \\
\partial_\nu F^{\nu\mu} &=& J^\mu, \\
\partial_\nu \hspace{0.3em} ^* F^{\nu\mu} &=& 0 .
\end{eqnarray}

The difference between equations (\ref{causalrmhdmass})--(\ref{causalrmhdampere}) 
and the RMHD equations used by the 
previous acausal resistive RMHD simulations \cite{watanabe06,komissarov07b} is
mainly in Ohm's law, as expected and suggested by other articles 
\cite{ardavan76,blackman93,gedalin96,melatos96,khanna98,meier04}.
The linear analysis of the electromagnetic wave in a pair plasma
shows that the inertia effect of the electron and
positron is essential in preserving causality.
Here, the electron/positron inertia term of Ohm's law is the right
hand side of equation (\ref{causalrmhdohm}).
If we neglect the change of $\gamma h /n$,
Ohm's law simplifies to
\begin{equation}
\VEC{E} + \VEC{V} \times \VEC{B} - \frac{\eta}{\gamma} [\VEC{J} - \gamma^2
(\rho_{\rm e} - \VEC{V} \cdot \VEC{J} ) (1+\Theta) \VEC{V}]
= \kappa \left [ 
\frac{\partial}{\partial t} (\VEC{J} + \rho_{\rm e} \VEC{V})  + \nabla \cdot \left
(\VEC{VJ} + \VEC{JV}
\right ) \right ], \label{causalrmhdohmm}
\end{equation}
where $\kappa = \tilde{h}/(2en)^2$.
When $H=\eta^2/\kappa < 1$, that is $\eta < \sqrt{\kappa}$, equations
(\ref{causalrmhdmass})--(\ref{causalrmhdenr}), 
(\ref{causalrmhddive})--(\ref{causalrmhdampere}), (\ref{causalrmhdohmm}) 
are causal, and thus we 
call equations (\ref{causalrmhdmass})--(\ref{causalrmhdenr}),
(\ref{causalrmhddive})--(\ref{causalrmhdampere}),(\ref{causalrmhdohmm}) 
with $\eta < \sqrt{\kappa}$ the ``causal
resistive RMHD" equations. 
Among the causal resistive RMHD equations, equation (\ref{causalrmhdohmm})
is most important; we call it the ``causal Ohm's law".
When we set $\Theta =0$ , equation (\ref{causalrmhdohmm}) reduces to
the simpler Ohm's law,
\begin{equation}
\VEC{E} + \VEC{V} \times \VEC{B} - \frac{\eta}{\gamma} \left [\VEC{J} - \gamma^2
(\rho_{\rm e} - \VEC{V} \cdot \VEC{J} ) \VEC{V} \right ]
= \kappa \left [ 
\frac{\partial}{\partial t} (\VEC{J} + \rho_{\rm e} \VEC{V})  + \nabla \cdot \left
(\VEC{VJ} + \VEC{JV}
\right ) \right ], \label{causalrmhdohmmm}
\end{equation}
which is quite similar to the generalized Ohm's law derived by 
\cite{ardavan76} and \cite{meier04}, but not identical.

\section{Expected Phenomena related to superluminal wave packet
\label{sec5}}

In this section, we discuss phenomena related to superluminal propagation 
of electromagnetic wave packets that appeared in RMHD simulations 
that use an acausal Ohm's law with $H=\eta^2/\kappa \agt 3$.
First, we show that it is difficult to detect the superluminal propagation of a 
electromagnetic wave packet in an unmagnetized plasma at rest.
For simplicity, we use the acausal Ohm's law
with $\kappa = 0$ ($H \rightarrow \infty$). 
This is just the case of the previous studies with
resistive RMHD \cite{watanabe06,komissarov07b}. 
In this case, the dispersion relation becomes that of the telegraphic
equation,
\begin{equation}
\omega^2 + \frac{i}{\eta} \omega - k^2 =0 .
\end{equation}
The group velocity of the dispersion relation 
$v_{\rm g} = \partial \omega/\partial k = k/[k^2 -(2\eta)^{-2}]^{1/2} > 1$ 
is always greater than the light speed in vacuum. The damping time of the wave is
$\tau_{\rm damp} = 1/(-\Im(\omega)) = 2\eta$. The diffusion time of the wave 
packet is calculated by
\[
\tau_{\rm diff} = \frac{\sigma^2}{|D|} 
= \frac{\sigma^2}{|\partial^2 \omega/\partial k^2|}
=\sigma^2 (2 \eta)^2 [k^2 - (2\eta)^{-2}]^{3/2},
\]
where $\sigma$ is the width of the wave packet (see Appendix \ref{appenc}, 
equation (\ref{decaytime})).
The life time of the wave packet is estimated by
\begin{equation}
\tau = \left ( 
\frac{1}{\tau_{\rm damp}} + \frac{1}{\tau_{\rm diff}} \right )^{-1}
=\frac{\sigma^2 (2 \eta)^2 [k^2 - (2\eta)^{-2}]^{3/2}}
{1 + \sigma^2 (2\eta) [k^2 - (2\eta)^{-2}]^{3/2}}.
\end{equation}
The characteristic propagation length of the wave packet is
\[
l = v_{\rm g} \tau  
=\frac{k \sigma^2 (2 \eta)^2 [k^2 - (2\eta)^{-2}]}
{1 + \sigma^2 (2\eta) [k^2 - (2\eta)^{-2}]^{3/2}}.
\]
Using $N=k\sigma$ and $\chi = 2 \eta k$, we have
\begin{equation}
\frac{l}{\sigma} = \frac{N\chi^2 (\chi^2 -1 )}{\chi^2 + N^2 (\chi^2 -1)^{3/2}}.
\end{equation}
Note that the limitation $l/\sigma \rightarrow \infty$ 
($\chi \rightarrow \infty$) means the wave packet propagates to a very long
distance compared to the scale of the wave packet itself in a highly
resistive plasma, where the situation is almost the same in a vacuum.
However, to detect the superluminal propagation of the wave packet, we have
to detect the difference between the propagation length of the wave packet 
and that of the light
in vacuum, $\Delta l = l - \tau$. The difference is estimated as
\begin{equation}
\frac{\Delta l}{\sigma} 
=\frac{l}{\sigma} \left ( 1 - \frac{1}{v_{\rm g}} \right )
= \frac{N(\chi^2 -1)}{[N^2 (\chi^2-1)^{3/2} + \chi^2] \left (1+\sqrt{1- \chi^{-2}}
\right )} \le \frac{2^{2/3}N}{3N^{4/3}+ 2^{2/3}}  .
\label{impdetsupl}
\end{equation}
Because $N \gg 1$, equation (\ref{impdetsupl}) shows $\Delta l \ll \sigma$.
This means detection of the superluminal propagation of the wave
packet is difficult in the rest background plasma with a detector of 
ordinary sensitivity.

\begin{figure}
\includegraphics[width=\textwidth]{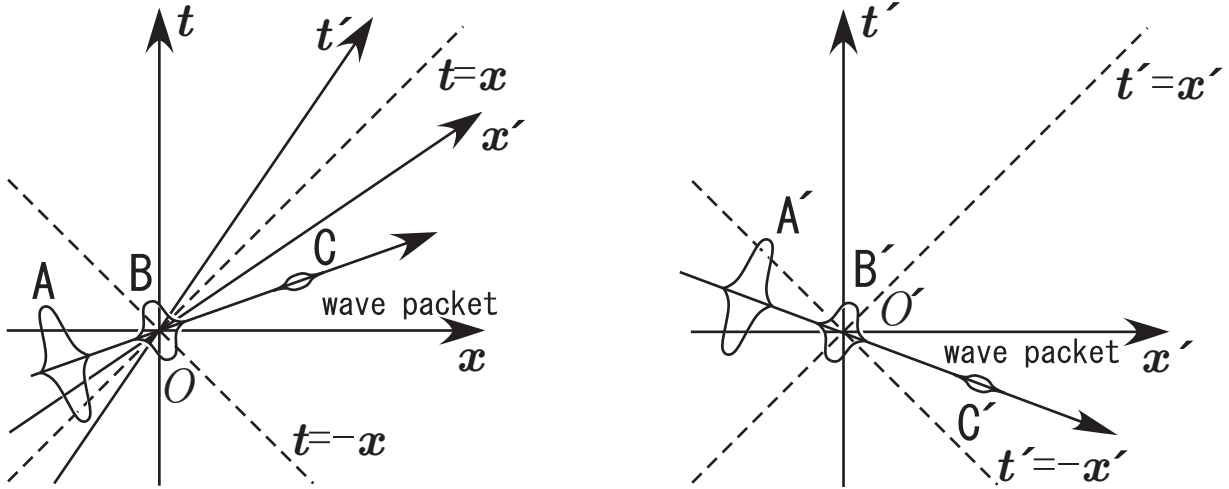}
\caption{Propagation of electromagnetic wave packet in
the uniform, unmagnetized plasma.
(a) Case of rest plasma. (b) Case of relativistic flow of plasma.
\label{funyphenom}
}
\end{figure}

When we consider a moving plasma with relativistic speed, 
propagation of the superluminal wave packet changes drastically.
Here we consider the wave packet propagating along the $x$ direction
of a frame $(t,x)$ in a
uniform, unmagnetized plasma at rest (see Fig. \ref{funyphenom}(a)).
We assume that the wave packet propagates with the group velocity $v_{\rm g} >1$
and damps with the damping rate $\gamma_{\rm dmp}$.
Next, we consider a new frame $(t',x')$ moving with velocity $v_0 > 1/v_{\rm g}$
relative to the frame $(t,x)$,
where the $t'$-axis and $x'$-axis in the space-time $(t,x)$ 
are drawn as shown in Fig. \ref{funyphenom}(a). The world line of the wave
packet is located between the $x'$-axis and $x$-axis. When we ride on the 
new frame $(t',x')$, we see from time inversion arguments 
that the wave packet propagates from 
the right to the left as shown in Fig. \ref{funyphenom}(b).
Furthermore, the wave packet grows at the rate 
$\gamma'_{\rm grw} = \gamma_{\rm dmp} \sqrt{1-v_0^2}/(v_0 v_{\rm g}-1)$.
Here the points A, B, and C with respect to the wave packet are identified
with those at A', B', and C', respectively.
%
This suggests that we have to use the causal RMHD equations 
(\ref{causalrmhdmass})--(\ref{causalrmhdenr}),
(\ref{causalrmhddive})--(\ref{causalrmhdampere}), and
(\ref{causalrmhdohmm}) 
to avoid such a strange instability of the wave packet, at least 
in relativistic plasma flow, because wave
packets propagating in such a flow will grow explosively.
Relativistic flow exists around the black hole horizon in
the Kerr space-time, so 
{\it artificial radiation of electromagnetic wave packets 
from the horizon will occur in acausal RMHD calculations}.
On the other hand, the same acausal RMHD equations with $\kappa = 0$ 
cause no problem
for a non-relativistically moving plasma.

\section{Concluding Remarks and Discussion
\label{sec6}}

We have derived the dispersion relation for electromagnetic waves
in a resistive pair plasma based on the relativistic two-fluid model, 
and have shown that the group velocity
of electromagnetic waves in a resistive plasma
is smaller than the light speed within the plasma
condition ($N_{\rm p} > 1$). 
This shows that superluminal communication is impossible
in a resistive plasma,
thus confirming the causal nature of signals in plasmas.
Furthermore, the causality condition, $H = \eta^2/\kappa< 2(3N_{\rm p})^{-2}<1$,
provides an upper limit for electric resistivity in resistive RMHD: 
\begin{equation}
\eta < \sqrt{\kappa} = \sqrt{\frac{h_0}{2mc^2n_0}} \frac{1}{\epsilon_0\omega_{\rm pe}}
= 0.2 \left ( \frac{n_{\rm 0}}{10^{20} {\rm m}^{-3}} \right )^{-1/2} 
\left ( \frac{h_0}{2mc^2n_0} \right )^{1/2}
\, [\Omega \, {\rm m}].
\end{equation}
For simplicity, we assumed that the relative velocity of the positron
and electron fluids is much smaller than their internal thermal velocities.
This is consistent with the assumption of linear analysis.
In general, however, this assumption is not valid, especially
for relativistic plasma around black holes.
To deal with plasmas where the relative velocity of the electron 
and positron fluids is relativistic, we have to return to the relativistic 
Vlasov--Boltzmann
equation with collisional terms \cite{ardavan76,blackman93,gedalin96,meier04}
to obtain the resistive term in the causal Ohm's law.
In such a case, resistivity depends on current density.

We emphasize that the inertia effect is important for preserving 
causality, i.e., to forbid the superluminal propagation of 
electromagnetic waves in a resistive plasma. 
If we neglect the inertia term in the 
generalized Ohm's law (the first term of the right hand side of
equation (\ref{rmhdohm})), 
then the resistive RMHD equations give a group velocity of electromagnetic
waves, $v_{\rm g} = \partial \omega/\partial k = 2k/(4k^2 - \eta^{-2})^{1/2}$,
that is greater than the light speed. This shows that the inertia 
effects of electrons and positrons should be considered to preserve causality.
We therefore proposed a set of causal
resistive RMHD equations in section \ref{sec4}.
Numerical techniques for simulating ``causal resistive
RMHD" flow should be developed quickly and be applied to 
astrophysical calculations ----- e.g., energy extraction from a rotating 
black hole by magnetic reconnection \cite{koide08}. 
To perform causal resistive RMHD simulations
of a black hole magnetosphere,
we are to use the general relativistic MHD equations along 
with the causal Ohm's law.

When we consider matter with a plasma parameter less
than one, we cannot use the simple two-fluid approximation,
because particles in the system are electrically bound to each other.
Metal, like iron, is an example for such matter.
To treat such a relativistic system, we ultimately must use 
relativistic quantum mechanics. However, we have no framework for that at present;
nevertheless it is an interesting and challenging field for future work.
In such an unknown framework, the group velocity of electromagnetic
waves in any medium should not be larger than the light speed 
(even if the wave damps quickly) to preserve causality. 
On the other hand, within the classical framework where we neglect 
quantum effects, we would show that the group velocity is always equal to or 
smaller than the speed of light when we treat the system properly. Here we cannot
use the method of smoothing the electromagnetic field, as is done in
traditional particle simulations, because the simple
smoothing destroys causality. Therefore, numerical calculations
may be more difficult than traditional plasma 
particle simulations.
It is interesting and important to investigate, 
in a medium with a small plasma parameter,
which effects (quantum or classical effects)
are more important for keeping the group velocity of electromagnetic wave
equal to or less than the speed of light.

In this paper, we considered only a pair plasma; 
we did not treat an electron--proton plasma. However, the similar conclusions
also should be drawn for the latter (at lease when the plasma is unmagnetized), 
because the linearized terms of resistive
RMHD in a pair plasma and in an electron--proton plasma
are expected to be similar.
It also is important to note the differences between
RMHD of a pair plasma and in an electron--proton plasma.
These come from the inequality between the mass ratios of 
the electron--positron and electron--proton.
With respect to equations (\ref{causalrmhdmass})--(\ref{causalrmhdenr}),
(\ref{causalrmhddive})--(\ref{causalrmhdampere}), (\ref{causalrmhdohmm}), 
it is expected that these equations are similar except for 
appearance of the second term in the brackets
on the left hand side of equation (\ref{causalrmhdmom}), $\VEC{JJ}/(2ne)^2$, and
the term 
in the brackets on the left hand side of equation (\ref{causalrmhdohmm}), 
$\gamma^2 (\rho_{\rm e} - \VEC{V} \cdot \VEC{J}) \Theta \VEC{V}$.
In the electron--proton plasma, the electron inertia term with $\VEC{JJ}$
is negligible compared to the proton inertia term with $\VEC{UU}$,
and $\Theta$ vanishes because of poor 
energy exchange between the electron and proton fluids. However, in
the pair plasma, $\Theta$ is not negligible.
Furthermore, the Hall effect disappears
in Ohm's law (\ref{causalrmhdohmm}) in the pair plasma case.
Note that all of the terms are nonlinear and that
the coefficient $\kappa$ of the inertia term
in the causal Ohm's law for the electron--proton plasma is
much smaller than that of the pair plasma
(by the ratio of the electron and proton masses).
It is believed that an accretion disk in a black hole magnetosphere of an AGN
will consist of an electron--proton plasma and a corona around 
the disk and a relativistic
jet from AGN consist of pair plasma \cite{wardle98}.
Comparison between phenomena in relativistic pair plasmas
and electron--proton plasmas is both interesting
and necessary for understanding the physics of black hole
magnetospheres where a relativistic jet may be produced. 



\begin{acknowledgments}
I thank Mika Koide, Takahiro Kudoh, 
Dongsu Ryu, Masaaki Takahashi, and Satoshi Yajima for this study. 
David L. Meier spent considerable effort checking my manuscript.
I appreciate his important comments and suggestions.
This work was supported in part by the Science Research Fund of the
Japanese Ministry of Education, Culture, Sports, Science and Technology.
\end{acknowledgments}

\appendix

\section{Derivation of frictional four-force density
\label{appena}}

In this appendix, we derive the friction four-force density between electron
and positron fluids, whose proper densities are $n_\pm$.
We use $f_-^\mu$ and $f_+^\mu$ to denote the friction density
of the electron and positron fluids, respectively.
The principle of action--reaction is expressed as
\begin{equation}
f_+^i + f_-^i = 0 \verb!   ! (i=1,2,3) ,
\end{equation}
in any inertial frame $x^\mu$. When we consider any other inertial
frame $X^\mu = A^\mu_\nu x^\nu$, the principle of action and reaction is
\begin{equation}
F_+^i + F_-^i = A^i_\nu (f_+^\nu + f_-^\nu) = A^i_0 (f_+^0 + f_-^0) = 0.
\end{equation}
Because $A^i_0 \neq 0$ in general, we have $f_+^0 + f_-^0 = 0$, or
\begin{equation}
f_+^\mu + f_-^\mu = 0  .
\label{actcoa}
\end{equation}
Note that $f_+^0 + f_-^0 =0$ is the law of conservation of energy.
We consider the center-of-mass frame of the two fluids 
$x^{\mu '} = a^\mu_\nu x^\nu$
where the four-velocity of the electron/positron fluids $U_\pm^{\mu '}$ satisfies
\begin{eqnarray}
n_+ U_+^{i '} + n_- U_-^{i '} = 0.
\label{cenofmas}
\end{eqnarray}
With respect to the inverse transformation, $x^\mu = b^\mu_\nu x^{\nu \prime}$,
we have
\begin{equation}
n_+ U_+^\mu + n_- U_-^\mu = b^\mu_\nu 
(n_+ U_+^{\nu \prime} + n_- U_-^{\nu \prime})
- b^\mu_0 (n_+ \gamma_+^\prime + n_- \gamma_-^\prime),
\end{equation}
where the prime denotes the variable observed in the center-of-mass frame.
Using the definition $U^\mu = (n_+ U_+^\mu + n_- U_-^\mu)/(2n)$ and
$\gamma = (n_+ \gamma_+ + n_- \gamma_-)/(2n)$, we have
\begin{equation}
b^\mu_0 = \frac{U^\mu}{\gamma'}.
\label{bm0}
\end{equation}
In the center-of-mass frame, the spacial components of the friction force density are
\begin{equation}
-f_-^{i '} = f_+^{i '} = 
-m\sigma_{\rm ee} v_{\rm r} n_+ n_- \gamma_+^\prime \gamma_-^\prime 
(v_+^{i \prime} -v_-^{i \prime}),
\end{equation}
where $\sigma_{\rm ee} $ is the electron/positron collisional cross section,
which is a function of the thermal velocity. 
The average relative velocity of the electrons and positrons, $v_{\rm r}$,
is roughly given by the maximum of the thermal velocity and the relative
velocity of the two fluids.
We write the friction four-force density as
\begin{eqnarray}
-f_-^\mu = f_+^\mu = b^\mu_\nu f_+^{\nu \prime}
&=& - m \sigma_{\rm ee}  v_{\rm r} n_+ n_- \gamma_+^\prime \gamma_-^\prime
(v_+^{i \prime} - v_-^{i \prime}) + b^\mu_0 f_+^{0 \prime} \nonumber \\
&=& - m \sigma_{\rm ee}  v_{\rm r} (n_-\gamma_-^\prime n_+ U_+^\mu 
- n_+ \gamma_+^\prime n_- U_-^\mu) + b^\mu_0 f_+^{0 \prime} .
\label{frc4frc1}
\end{eqnarray}
When we use the collision frequency of the electron and positron
$\nu_{\rm ee} = \sigma_{\rm ee}  v_{\rm r} n$ and equation (\ref{bm0}), we have
\begin{equation}
f_+^\mu = - \frac{m\nu_{\rm ee}}{n} 
(n_- \gamma_-^\prime n_+ U^\mu_+ - n_+ \gamma_+^\prime n_- U^\mu_-)
+ \frac{n_+ U^\mu_+ + n_- U^\mu_-}{n_+ \gamma_+^\prime + n_- \gamma_-^\prime} f_+^{0 \prime} .
\label{frc4frc2}
\end{equation}

Next we consider the energy gain rate of the positron fluid $f_+^{0\prime}$
in the center-of-mass frame. The positron and electron fluids lose the 
kinetic energy due to friction at the rate,
\begin{eqnarray}
-f_+^{i \prime} v_{+ i}' - f_-^{i \prime} v_{- i}' 
&=& - f_+^{i \prime} v_{+ i}' - \frac{1}{n_- \gamma_-'} f_-^{i \prime} n_- U_{- i}' \nonumber \\
&=& - f_+^{i \prime} v_{+ i}' - \frac{1}{n_- \gamma_-'} f_+^{i \prime} n_+ U_{+ i}'
= - f_+^{i \prime} v_{+ i}' \left ( 1 + \frac{n_+ \gamma_+'}{n_- \gamma_-'} \right ).
\end{eqnarray}
In the above calculation, we employ the principle of action--reaction 
(\ref{actcoa}), the condition of the center-of-mass frame (\ref{cenofmas}),
and the assumption that
the lost energy is thermalized. 
A fraction $\theta$ of this thermalized energy 
($0 \le \theta \le 1$) is distributed to the positron and electron fluids,
assuming equipartition, and other part is returned
to the original fluid. Then the energy gain rate of the positron
is calculated as
\begin{eqnarray}
f_+^{0 \prime} &=& \frac{- \theta f_+^{i \prime} v_{+ i}'
- \theta f_-^{i \prime} v_{- i}'}{\gamma_+' n_+ + \gamma_-' n_-}
\gamma_+' n_+ 
+(1-\theta) (- f_+^{i \prime} v_{+ i}') - (- f_+^{i \prime} v_{+ i}) \\
&=& - \theta  
f_+^{i \prime} v_{+ i} \frac{\gamma_+' n_+ - \gamma_- n_-}{n_- \gamma_-'} .
\label{enegainposi1}
\end{eqnarray}
Using the definition of the average four-velocity and four-current density
(\ref{ave4vel}), (\ref{ave4cur}) and the friction force density expression 
(\ref{frc4frc2}), 
we have
\begin{eqnarray}
f_+^{i \prime} v_{+ i}' &=& - m \sigma_{\rm ee} v_{\rm r} n_+ n_- \gamma_+' \gamma_-'
(v_+^{i \prime} - v_-^{i \prime}) v_{+ i}' \nonumber \\
&=& - \frac{m \sigma_{\rm ee} v_{\rm r} n}{2 n_+ \gamma_+' \gamma' e^2} [Q^2 - WM] ,
\label{fidpvdpi}
\end{eqnarray}
where $M=U_\nu U^\nu$, $Q=J_\nu U^\nu$, $W=J_\nu J^\nu$.
Equations (\ref{enegainposi1}) and (\ref{fidpvdpi}) yield
\begin{eqnarray}
f_+^{0 \prime} &=& 
 \frac{m \sigma_{\rm ee} v_{\rm r} n}{2 e^2 n_+ \gamma_+' \gamma' n_- \gamma_-'}
(Q^2 - WM) (\gamma_+' n_+ - \gamma_-' n_-) \theta \nonumber \\
&=& - \frac{m \sigma_{\rm ee} v_{\rm r} n^2/e^2}{(n^2M)^2- \left ( \frac{n}{2e}Q \right )^2}
(Q^2 - WM) \frac{nQ}{2e} \theta.
\label{eneexc}
\end{eqnarray}
Here we used
\begin{eqnarray}
\gamma_+' n_+ - \gamma_-n_- ' &=& \gamma' J^{0 \prime}
= - J^{0 \prime} U_0' = - J^{\nu \prime} U_\nu' = - J^\nu U_\nu = -Q , \nonumber \\
\gamma'^2 &=& - U^{0 \prime} U_0' = - U^{\nu \prime} U_\nu' 
= - U^\nu U_\nu = -M , \nonumber \\
\gamma'^2 n_+  \gamma_+' n_- \gamma_-' &=& 
\gamma'^2 \left ( nU^{0 \prime} + \frac{J^{0 \prime}}{2e} \right )
\left ( nU^{0 \prime} - \frac{J^{0 \prime}}{2e} \right ) \nonumber \\
&=& \frac{1}{n^2} \left [
(n^2 U_\nu U^\nu)^2 - \left ( \frac{n}{2e} J_\nu U^\nu \right )^2
 \right ]  .
\end{eqnarray}
The equation of friction four-force density (\ref{frc4frc2}) reads 
\begin{eqnarray}
f_+^\mu &=& - m \sigma_{\rm ee} v_{\rm r} \left [
n_- \gamma_-^\prime \left ( n U^\mu + \frac{1}{2e} J^\mu \right )
- n_+ \gamma_+^\prime \left ( n U^\mu - \frac{1}{2e} J^\mu \right )
\right ] + \frac{U^\mu}{\gamma'} f_+^{0 \prime} \\
&=& - \frac{m\sigma_{\rm ee} v_{\rm r}n}{\gamma' e}
\left [ -(U_\nu U^\nu) J^\mu + (U_\nu J^\nu) U^\mu \right ]
+ \frac{U^\mu}{\gamma'} f_+^{0 \prime} .
\end{eqnarray}
When we introduce the dimensionless factor with respect to the
left hand side of equation (\ref{eneexc}),
\begin{equation}
\Theta = \frac{2m^2}{e^2} \frac{Q^2-MW}{(2mnM)^2-(mQ/e)^2} \theta  ,
\end{equation}
we finally obtain
\begin{equation}
f_+^\mu = - \frac{n \sigma_{\rm ee}  v_{\rm r} n}{e} \sqrt{-M}
\left [ J^\mu - \frac{Q}{M} U^\mu (1+\Theta) \right ].
\end{equation}
We also calculate the resistive term in Ohm's law,
\begin{equation}
\frac{f_+^\mu}{en\gamma} = - \eta \frac{\sqrt{-M}}{\gamma}
\left [ J^\mu - \frac{Q}{M} U^\mu (1+\Theta) \right ],
\end{equation}
where $\eta \equiv m \sigma_{\rm ee}  v_{\rm r}/e^2$ is resistivity.
When we use the collision frequency $\nu_{\rm ee} = \sigma_{\rm ee}  v_{\rm r} n$,
we can write $\eta = m \nu_{\rm ee}/(ne^2)$.


\section{Forbidden range of superluminal communication
\label{appenb}}

We prove that $\partial \Omega/\partial k < 1$ in the dispersion 
relation (\ref{disprealomega}) when $H < 1.5$.
In this appendix, we omit the hat over $\hat{k}$.
The determinant of the cubic equation (\ref{disprealgamma}) with
respect to $\gamma$ is $D_\gamma = -(C^3+F^2)/(9 \times 36^2)$.
If $\gamma$ had three different real solutions, it would yield $D_\gamma > 0$, i.e.,
$\Omega^2 (4\Omega^2 -C)^2 = (C^3+F^2)/27<0$ 
($\Omega$ would be pure imaginary).
Then $\gamma$ has only one real solution, so that $\Omega$ has a real solution.
When we consider a function of the left hand side of equation (\ref{disprealgamma}),
\[
f(\gamma) = \gamma^3 - \gamma^2 - \frac{1}{4} \left (
1 + k^2 + \frac{1}{H} \right )\gamma - \frac{1}{8H},
\]
we have $f(0) = -1/(8H)<0$ and $f(1/2)=k^2/8>0$, and then
the single solution of $\gamma$ should be $0< \gamma < 1/2$.
Then we have
\begin{equation}
k^2 + \frac{1}{H} - \frac{1}{3} < \Omega^2 < k^2 + \frac{1}{H}  ,
\label{rangeomega2}
\end{equation}
because
\begin{equation}
\Omega^2 = 3\gamma ( \gamma - \frac{2}{3}) + k^2 + \frac{1}{H}
=3 \left ( \gamma - \frac{1}{3} \right )^2 - \frac{1}{3} + k^2 + \frac{1}{H}.
\label{omega2gamma}
\end{equation}
From equation (\ref{omega2gamma0}), we have
\begin{equation}
\Omega \frac{d \Omega}{dk} = (3\gamma -1) \frac{d \gamma}{dk} + k.
\label{domegadk}
\end{equation}
Using equation (\ref{disprealgamma}), we obtain
\begin{equation}
\frac{d \gamma}{dk} = k \left [ 1 - \frac{1}{2} 
\frac{3\gamma^2 -\gamma}{3 \gamma^2 - 2\gamma + \frac{1}{4} (1 + k^2 + 1/H)}
\right ].
\label{dgammadk}
\end{equation}
When $H < 3$, the denominator in equation (\ref{dgammadk}) is positive because of
the right side equation of equation (\ref{omega2gamma}).
From equations (\ref{domegadk}) and (\ref{dgammadk}), we have
\begin{equation}
\frac{d\Omega}{dk} = \frac{k}{2\Omega}
\frac{\Omega^2 - \frac{k^2}{2} - \frac{1}{2H} + \frac{1}{2} -\gamma }
{\Omega^2 - \frac{3k^2}{4} - \frac{3}{4H} + \frac{1}{4} }   .
\label{domegadk2}
\end{equation}
We consider the difference between the numerator and positive denominator
of equation (\ref{domegadk2}),
\begin{equation}
\Delta = k \left (\Omega^2 - \frac{k^2}{2} - \frac{1}{2H} + \frac{1}{2} -\gamma \right )
-2 \Omega \left  (\Omega^2 - \frac{3k^2}{4}) -\frac{3}{4H} + \frac{1}{4} \right )  .
\end{equation}
After some algebraic calculations, we have
\begin{equation}
\Delta = (k-\Omega) \left ( \Omega^2 -k^2 - \frac{1}{2H} \right )
-\frac{\Omega}{4} \left ( 4 \Omega^2 - 4k^2 - \frac{4}{H} +2 \right )
+ \left ( \frac{1}{2} - \gamma \right ) k
- \frac{3}{4} \Omega k^2 + (k-\Omega) \frac{k^2}{2}.
\end{equation}
From the left inequality in equation (\ref{rangeomega2}), we find
$\Omega > k$ when $H<3$. And then we have
\begin{equation}
\Delta \le (k-\Omega) \left ( \frac{1}{2H} - \frac{1}{3} \right )
+ (k-\Omega) \frac{k^2}{2} - \frac{\Omega}{6} 
+ \left ( \frac{1}{2} - \gamma \right ) k - \frac{3}{4} k^2.
\end{equation}
Using equation (\ref{omega2gamma0}), we obtain
\begin{equation}
\Delta \le (k-\Omega) \left ( \frac{1}{2H} - \frac{1}{3} \right )
+\frac{k-\Omega}{6} + \frac{k}{2} \left [
\frac{2}{3} - \frac{1}{H} + \Omega \left ( \Omega - \frac{5k}{2} \right)
-3 \gamma^2 \right ].
\end{equation}
When $k \ge 2/\sqrt{21H}$, we get $\Omega^2 < k^2 + 1/H \le 25k^2/4$
using equation (\ref{rangeomega2}). Then we have $\Omega \ge 5k/2$ and
$\Delta < 0$ when $H<3/2$.

On the other hand, when $k < 2/\sqrt{21H}$, we obtain
 $\Omega^2 < k^2 + 1/H < 25/(21H)$ and then $\Omega < 5/\sqrt{21H}$.
After some calculations, we have
\begin{eqnarray}
\Delta & \le &  (\Omega -k) \left ( 
\frac{1}{6} -\frac{1}{2H} + \frac{k\Omega}{2} -\frac{3k^2}{4} \right )
+ \frac{k}{2} \left ( \frac{2}{3} - \frac{1}{H} \right )
-\frac{3k^2}{4} - \frac{3\gamma^2 k}{2} \nonumber \\
& < & (\Omega -k) \left ( 
\frac{1}{6} -\frac{11}{42H} -\frac{3k^2}{4} \right )
+ \frac{k}{2} \left ( \frac{2}{3} - \frac{1}{H} \right ) 
-\frac{3k^2}{4} - \frac{3\gamma^2 k}{2} < 0   ,
\end{eqnarray}
when $H \le 3/2$. Summarizing above calculations, we conclude that
$\Delta < 0$ when $H \le 3/2$. This shows that 
$v_{\rm g} = \partial \Omega/\partial k < 1$ when $H<3/2$.

\section{Propagation and damping of electromagnetic wave packets
\label{appenc}}

Here we consider the propagation of a packet of electromagnetic
waves in a resistive plasma.
The wave packet is regarded as an element for communication in the medium.
First, we use the analytic approximation of a wave packet with a large width.

\subsection{An analytic approximation solution}

Any variable perturbation of the electromagnetic wave packet 
in resistive pair plasma, $f_1$, is given by,
\begin{equation}
f_1 = \int_{-\infty} ^\infty F(k) e^{ikx - i\omega(k) t} dk  ,
\label{analyticsolution0}
\end{equation}
where $F(k)$ is the Fourier transformation of the variable $f_1$. 
We take the Gaussian distribution of the wave packet to be
\begin{equation}
F(k) \propto \frac{\sigma}{\sqrt{2\pi}} e^{-\frac{\sigma^2}{2} (k-k_0)^2}  ,
\label{gaussdistr}
\end{equation}
where $\sigma$ is the width of the wave packet and $k_0$ is 
the characteristic wave number.
The initial profile of the variable $f_1$ is proportional to
$\exp[-x^2/(2\sigma^2)] \exp(ik_0x)$.

When $1/\sigma$ is much smaller than the characteristic scale $\Delta k$
of the dispersion relation with respect to $\omega(k)$, we use an approximation
\begin{equation}
f_1 = \int_{k_0-}^{k_0+} F(k) e^{ikx - i\omega(k)t} dk
\approx \int_{k_0-}^{k_0+} F(k) e^{ikx - i\left [ 
\omega(k_0) + \frac{\partial \omega}{\partial k} (k_0) (k - k_0) 
+ \frac{1}{2} \frac{\partial^2 \omega}{\partial k^2} (k_0) (k - k_0)^2
\right ] t} dk ,
\end{equation}
where we have expanded to 2nd order in $k - k_0$.
When we write $u=\partial k / \partial \omega (k_0)$,
$D=\partial^2 \omega / \partial k^2 (k_0)$,
\begin{eqnarray}
f_1 \propto \frac{\sigma}{\sqrt{2\pi}} e^{ik_0x - i \omega(k_0) t}
\int_{0-}^{0+} e^{-\frac{1}{2} (\sigma^2 + iDt) k'^2 
+ i(x-ut)k' } dk' \\
 \approx \frac{\sigma}{\sqrt{2\pi}} e^{ik_0x - i \omega(k_0) t}
\int_{-\infty}^{+\infty} e^{-\frac{1}{2} (\sigma^2 + iDt) k'^2 
+ i(x-ut)k' } dk' \\
= \frac{\sigma}{\sqrt{2\pi}} e^{ik_0x - i \omega(k_0) t}
\sqrt{\frac{\pi}{|\sigma^2+iDt|^2+\Re({\sigma^2+iDt})}}
\left ( 1 + \frac{\sigma^2-iD^*t}{|\sigma^2+iDt|^2} \right ) \nonumber \\
\times \exp \left [-\frac{1}{2} (\sigma^2 + iDt) (x-ut)^2 \right ]  .
\end{eqnarray}
Here we used a formula,
\begin{equation}
\int_{-\infty}^\infty e^{-\frac{c}{2}x^2} dx
= \sqrt{\frac{2\pi}{c}}
= \sqrt{\frac{\pi}{|c| + \Re{c}}} \left ( 1+ \frac{c^*}{|c|} \right )
\verb!   ! (c \in {\bf R})  ,
\end{equation}
where $c$ is an arbitrary complex constant.
When we write $u=u_{\rm r} + i u_{\rm i}$ and $D=D_{\rm r} + i D_{\rm i}$
($u_r, u_i, D_r, D_i \in \VEC{R}$),
after some algebraic calculations, we have 
\begin{eqnarray}
f_1 &= \frac{1}{\sqrt{2}} 
\left [ 1 + \frac{\sigma^2-iD_{\rm i}t -iD_{\rm r}t}
{\sqrt{(\sigma^2 - D_{\rm i} t)^2+(D_{\rm r}t)^2}} \right ] 
\frac{1}{\sqrt{\sqrt{(1-D_{\rm i}t/\sigma^2)^2+(D_{\rm r}t/\sigma^2)^2}
+1-D_{\rm i}t/\sigma^2}}  \\
& \times \exp \left [ ik_0 x - i\omega(k_0) t 
-i \frac{2(x-u_{\rm r}t)u_{\rm i}(\sigma^2-D_{\rm i}t )t + 
D_{\rm r} \{ (x-u_{\rm r}t)^2-(u_{\rm i}t)^2 \} t}
{2\{ (\sigma^2 -D_{\rm i} t)^2 + (D_{\rm r} t)^2 \}} \right ] \\
& \times \exp \left [ - \frac
{(\sigma^2 -D_{\rm i}t) \left ( x - u_{\rm r} t 
- \frac{D_{\rm r} u_{\rm i}}{\sigma^2 - D_{\rm i} t} t^2 \right )^2
-\frac{(D_{\rm r} u_{\rm i})^2}{\sigma^2 - D_{\rm i} t} t^4
-u_{\rm i}^2 (\sigma^2 - D_{\rm i} t)t^2}
{2\{ (\sigma^2 - D_{\rm i}t)^2 + (D_{\rm r}t)^2) \}}
\right ]  .
\end{eqnarray}
The width of the wave packet is approximately given by 
\begin{equation}
\Delta x \sim \sqrt{\frac{(\sigma^2 - D_{\rm i}t)^2+(D_{\rm r}t)^2}
{\sigma^2-D_{\rm i}t}}.
\end{equation}
When $t \ll \frac{u_{\rm r} \sigma^2}{u_{\rm r} D_{\rm i} + D_{\rm r} u_{\rm i}}$,
the propagation velocity of the wave packet is
\begin{equation}
v_{\rm packet} \sim u_{\rm r}.
\end{equation}
The diffusion time scale of the wave packet, $T_{\rm D}$, 
is found from the condition
$(\Delta x)^2 = 2 \sigma^2$, which yields
\begin{equation}
T_{\rm D} = \frac{\sigma^2}{\sqrt{D_{\rm r}^2 + D_{\rm i}^2}}
=\frac{\sigma^2}{|D|}  .
\label{decaytime}
\end{equation}
The propagation velocity, $v_{\rm packet}$, has meaning
only when $t \ll T_{\rm D}$ or $\sigma \gg \sqrt{t|D|}$.

\subsection{A method of numerical integration and application for imaginary
superluminal propagation of wave packet}

We show a numerical solution for an electromagnetic wave packet
propagation in resistive pair plasma, using Simpson's formula.
The variable perturbation of the electromagnetic wave packet $f_1$ given by 
equations (\ref{analyticsolution0}) and (\ref{gaussdistr}) is calculated as
\begin{equation}
\Re(f_1) \propto e^{-\gamma_0 t} P(x,t)  ,
\end{equation}
\begin{equation}
P(x,t) = \frac{\sigma}{\sqrt{2\pi}}
\int_{k_0-M/\sigma}^{k_0+M/\sigma}
e^{-\frac{\sigma^2}{2} (k-k_0)^2 - (\gamma(k) - \gamma_0) t}
\cos(kx-\Omega(k)t) dk  ,
\end{equation}
where $\gamma_0$ is the characteristic damping rate of the wave packet.
When the number $M$ is large enough, the integration becomes the exact value.
Usually we set $M=$4--10, which gives precise enough evaluation.
To calculate the profile of the wave packet, we evaluate the profile function
$P(x,t)$.

\begin{figure}
\includegraphics[width=0.9\textwidth]{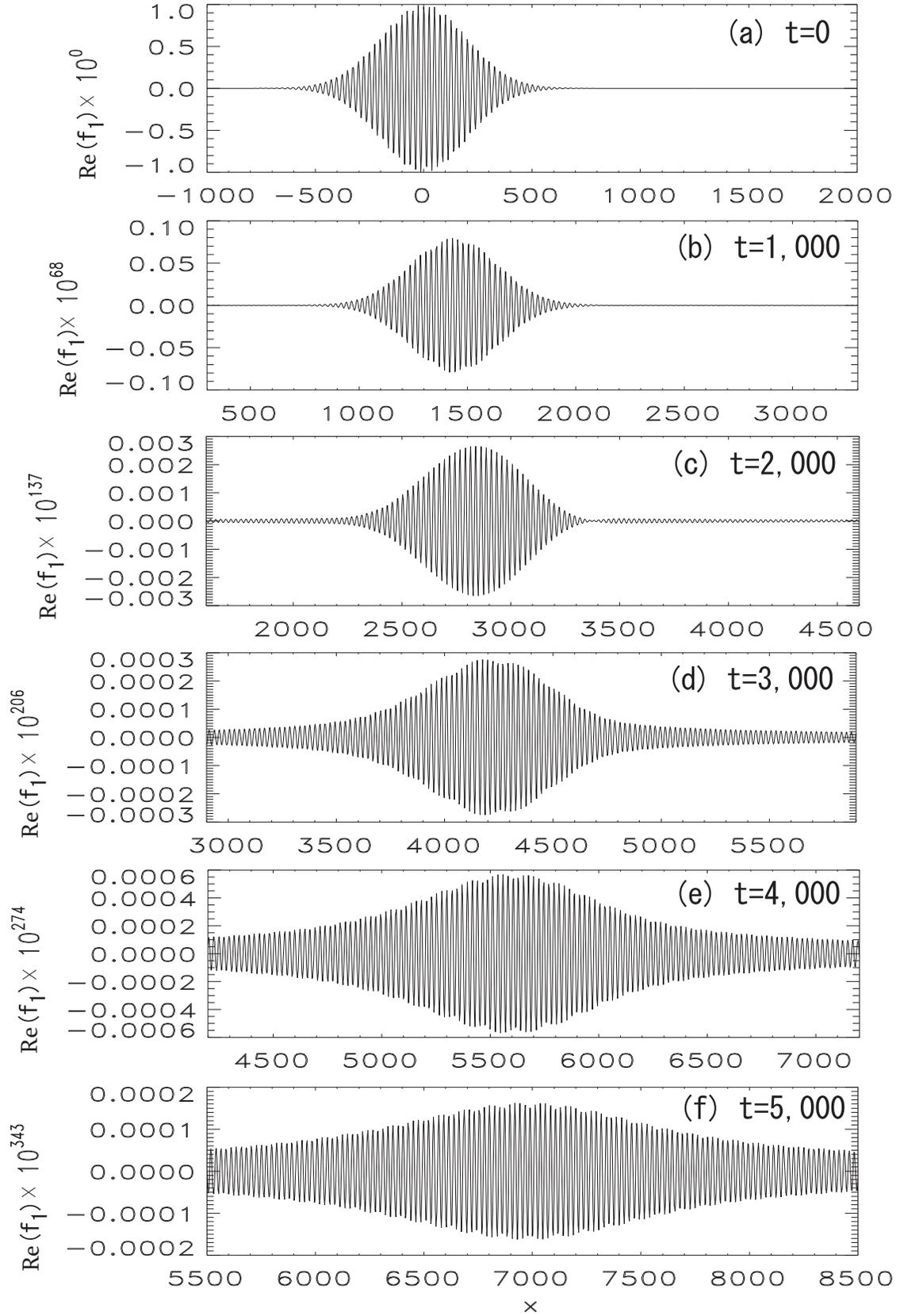}
\caption{
\label{timevol}
Imaginary time evolution of wave packet superluminal propagation 
in pair plasma, with $\sigma = 50$, $k_0=0.25$, $H=4$.
Here $f_1$ is normalized by the maximum initial value.
}
\end{figure}

Figure \ref{timevol} shows mathematically the imaginary time evolution 
of a superluminal electromagnetic wave packet with
$\sigma=50$, $k_0=0.25$ in the pair plasma when $H=4$ in equation (\ref{dispcomlex}).
The magnification rate of the variable $f_1$ is indicated by the factor
beside the ordinate.
The propagation velocity of the wave packet is found to be $v_{\rm packet} \sim 1.4$,
i.e., superluminal. The group velocity $u_{\rm r}=1.47$ gives a good approximation
to the propagation velocity of the wave packet.
When $t=4,000 \alt T_{\rm D} = 5,000$, the wave packet
begins diffuse. The value of $T_{\rm D}$ gives a good estimate of 
wave packet break-down.
This numerical calculation clearly shows that superluminal 
propagation of a wave packet
is possible if the group velocity of electromagnetic wave 
exceeds the speed of light.


\bibliography{apssamp}

\end{document}